\documentclass[aps,twocolumn,groupedaddress,floatfix,superscriptaddress]{revtex4}

\begin{document}
\bibliographystyle{apsrev}

\title{Tensor renormalization group approach to 2D classical lattice models}

\author{Michael Levin}
\affiliation{Department of Physics, Harvard University, Cambridge, Massachusetts 
02138}

\author{Cody P. Nave}
\affiliation{Department of Physics, Massachusetts Institute of Technology, Cambridge, Massachusetts 02139}

\begin{abstract}
We describe a simple real space renormalization group technique for two dimensional classical lattice models. The approach is similar in spirit to block spin methods, but at the same time it is fundamentally based on the theory of quantum entanglement. In this sense, the technique can be thought of as a classical analogue of DMRG. We demonstrate the method - which we call the tensor renormalization group method - by computing the magnetization of the triangular lattice Ising model. 

\end{abstract}

\maketitle

\textsl{Introduction}:
The density matrix renormalization group (DMRG) technique has proved extraordinarily powerful in the analysis of one dimensional quantum systems. \cite{W9263,S0559} Thus it is natural to try to develop an analogous renormalization group method in higher dimensions. Such a method could solve many currently intractable problems (such as the 2D Hubbard model).

Recent work has focused on generalizing DMRG to higher dimensional quantum systems. \cite{VC0466} But it is also natural to try to generalize to higher dimensional \emph{classical} lattice models. While classical real space renormalization group methods (such as block spin methods \cite{W7573}) have been around for many years, they have never achieved the generality or precision of DMRG.

In this paper, we address this problem in the two dimensional case. We use ideas from quantum information theory to develop a numerical renormalization group method that can effectively solve any two dimensional classical lattice model. The technique - which we call the tensor renormalization group (TRG) method - has no sign problem and works equally well for models with complex weights. 

Accurate numerical methods based on transfer matrices \cite{N9598,MVC0506} have already been developed for 2D classical systems. The advantage of the approach described here is that it is a \emph{fully isotropic} coarse graining procedure, similar in spirit to block spin methods. It is thus naturally suited to investigating universal long distance physics. Also, on a more theoretical level, the method reveals the relationship between classical RG and quantum entanglement. Finally, if only for its simplicity, we feel that the method is a useful numerical tool in two dimensions as well as a natural candidate for higher dimensional generalizations.

\textsl{Tensor network models}:
The tensor renormalization group method applies to a set of classical lattice models
called "tensor network models." \cite{SDV0570} Many well known statistical mechanical models, such as the Ising model, Potts 
model, and the six vertex model, can be written naturally as tensor network models. In fact, as we show later, all classical 
lattice models with local interactions can be written as tensor network models. 

\begin{figure}[tb] 
\centerline{
\includegraphics[width=1.1in]{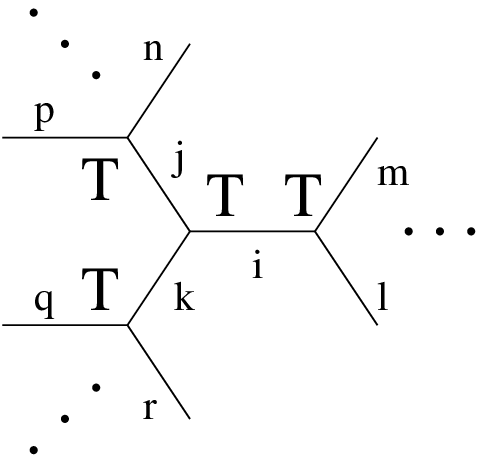}
}
\caption{
A tensor network model on the honeycomb lattice.
}
\label{Hlatt4}   
\end{figure}

To describe a tensor network model on the honeycomb lattice, one must specify a 
(cyclically symmetric) tensor $T_{ijk}$ with indices $i,j,k$ running from $1$ to $D$ 
for some $D$. The corresponding tensor network model has a degree of freedom $i=1,...,D$ on each bond of the honeycomb lattice. The weight for a configuration $(i,j,k,...)$ is given by
\begin{equation}
e^{-S(i,j,k,...)} = T_{ijk} T_{ilm} T_{jnp} T_{kqr}...
\end{equation}
where the product includes a tensor for each site of the lattice (Fig. \ref{Hlatt4}). 
The partition function is the sum of all the weights:
\begin{equation}
Z = \sum_{ijk...} e^{-S(i,j,k,...)} = \sum_{ijk...}T_{ijk} T_{ilm} T_{jnp} T_{kqr}...
\label{parttensor}
\end{equation}
In other words, the partition function is obtained by taking the product of all the 
tensors, contracting the pairs of indices on each bond. 

\begin{figure}[tb] 
\centerline{
\includegraphics[width=3.0in]{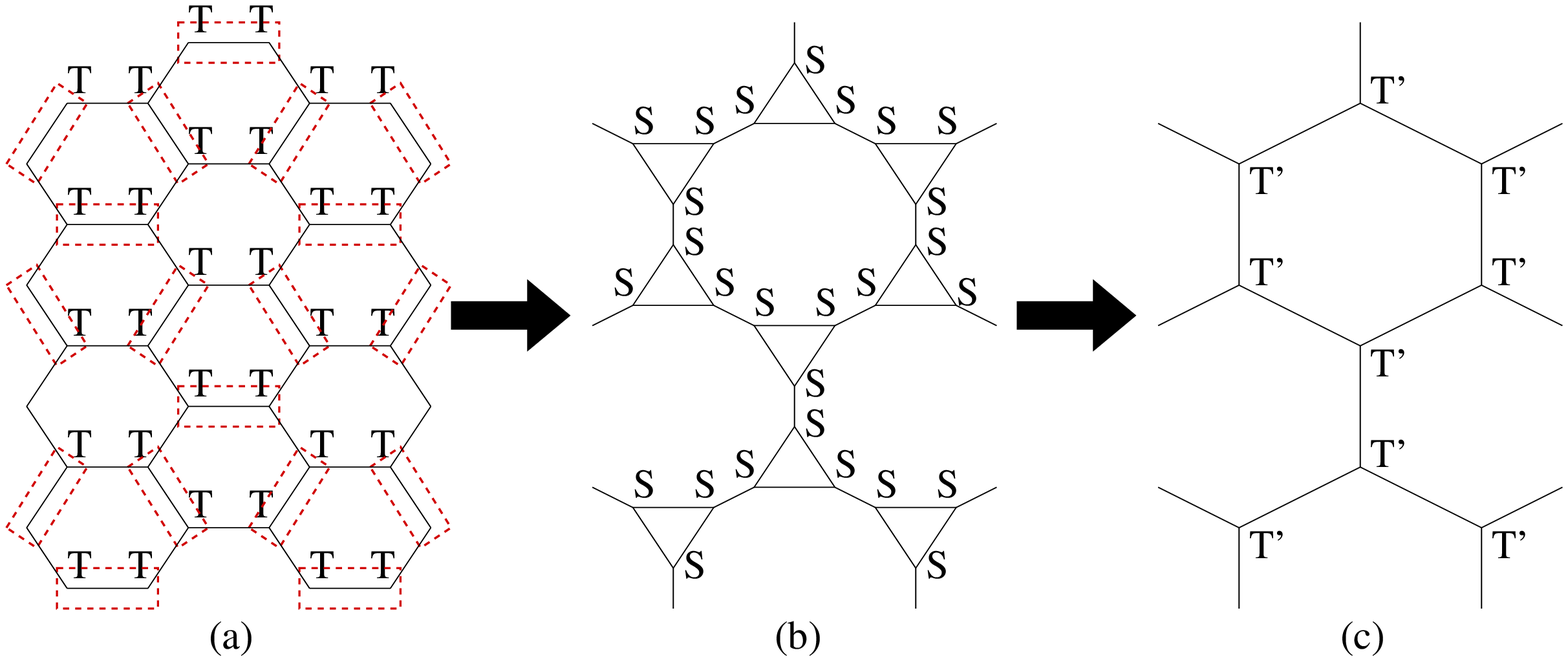}
}
\caption{
A TRG transformation on the honeycomb lattice. 
}
\label{rgsum}   
\end{figure}

\textsl{The TRG method}:
The tensor renormalization group method is a way to compute the partition function $Z$ using a real space RG flow. We explain the method in the case of the honeycomb lattice.
Each coarse graining iteration is made up of two separate steps. The first is approximate, and the second is exact (Fig. \ref{rgsum}). The first step is to find a tensor $S$ such that
\begin{equation}
\sum_{n} S_{lin} S_{jkn} \approx \sum_{m} T_{ijm} T_{klm}
\label{reconneq}
\end{equation}
(We explain how to find such a tensor $S$ at the end of this section). The relation 
(\ref{reconneq}) is a very useful property. Geometrically, it means that we can 
reconnect the lattice, making the replacement
\begin{align}
\includegraphics[height=0.35in]{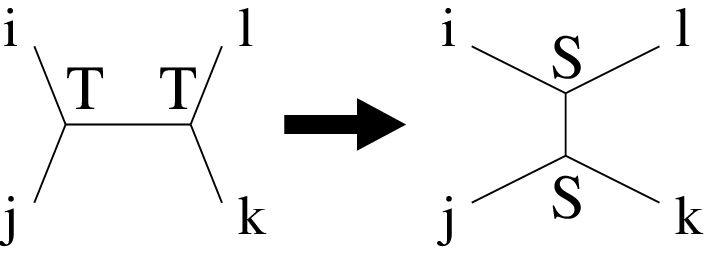} \label{reconn}
\end{align}
wherever we like, without affecting the partition function $Z$. Applying this to the bonds shown in Fig. \ref{rgsum}a, we change the honeycomb lattice to the new lattice shown in Fig. \ref{rgsum}b. The partition function is now given by contracting the $S$ tensors on the new lattice.

The second step is now clear. We group together triplets of neighboring points replacing
them by a single lattice point with a coarse-grained tensor $T'$:
\begin{align}
\includegraphics[height=0.35in]{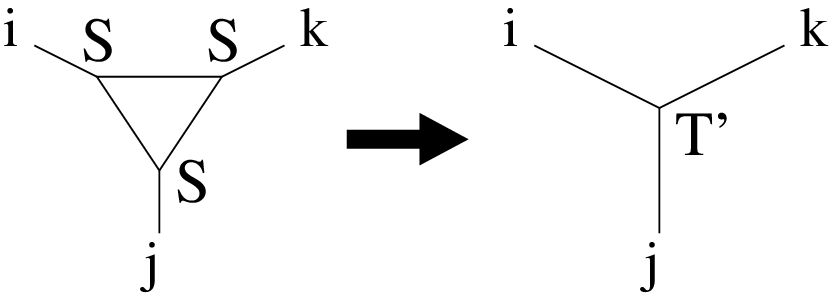}
\label{coarsegrain}
\end{align}
Here the tensor $T'$ is given by contracting over the three bonds of the triangle:
\begin{equation}
T'_{ijk} = \sum_{pqr} S_{kpq}S_{jqr}S_{irp}
\end{equation}

Making this replacement everywhere gives a new (coarser) honeycomb lattice (see Fig. \ref{rgsum}c). This completes the coarse graining transformation. The end result is that the number of points in the lattice has decreased by a factor of $3$ and $T$ has been replaced by $T'$. 

\begin{figure}[tb] 
\centerline{
\includegraphics[width=3.0in]{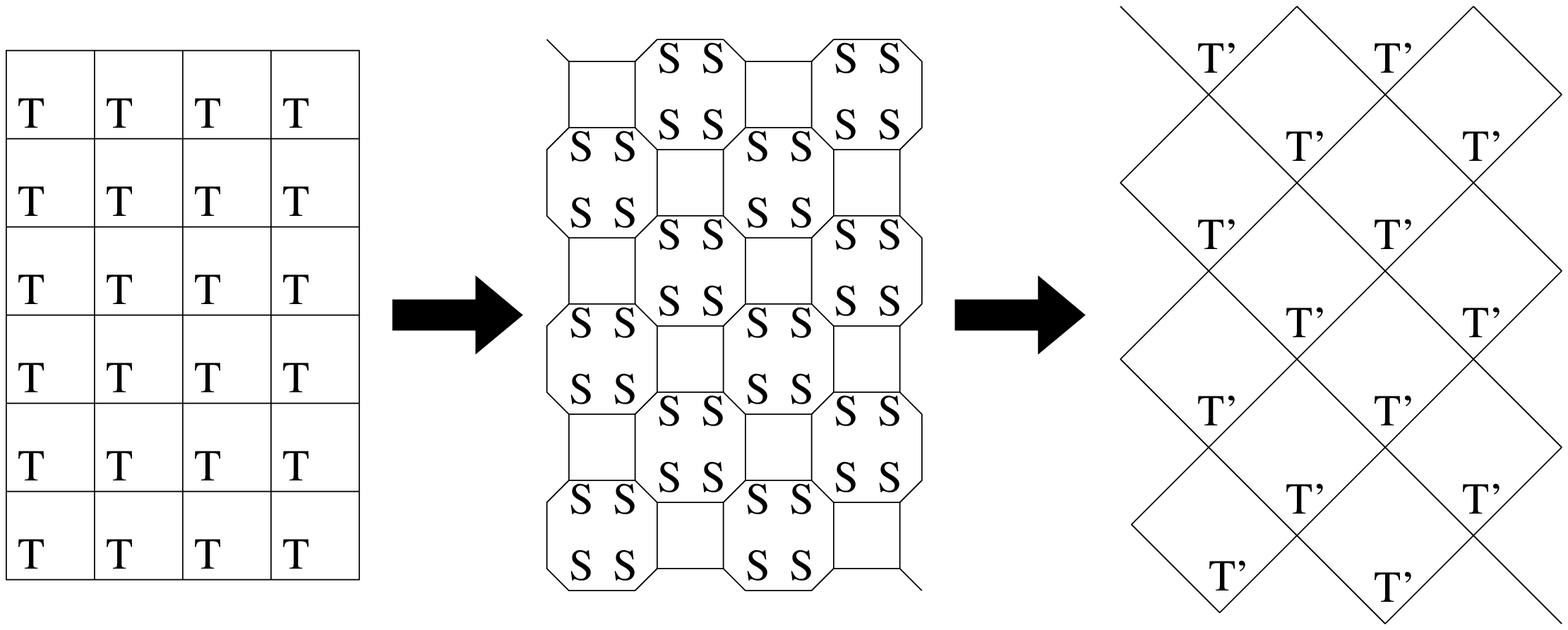}
}
\caption{
A TRG transformation on the square lattice. 
}
\label{rgsumsq}   
\end{figure}

Iterating this procedure, one can compute the partition function of an 
arbitrarily large finite lattice. Thermodynamic observables and correlation functions can be obtained by taking numerical derivatives of $F = -\log Z$, or by evaluating the free energy of more general models where the tensors $T_{ijk}$ vary from site to site. An additional feature is that the tensors $T$ converge to a fixed point tensor $T^*$ - whose physical significance we explain in the next section. The method is not limited to the honeycomb lattice and can easily be implemented on other lattices (see Fig. \ref{rgsumsq}).

To complete the discussion, we address the issue of finding a tensor
$S$ which satisfies (\ref{reconneq}). The first step is to think of $\sum_{m} 
T_{ijm} T_{klm}$ as a $D^2 \times D^2$ matrix $M$: $M_{li,jk} = \sum_{m} T_{ijm} T_{klm}$.
It is also useful to think of the tensor $S_{lin}$ as a $D^2 \times D$ 
matrix $S_{li,n}$. Then the problem of satisfying (\ref{reconneq}) 
is the problem of finding a $D^2 \times D$ matrix $S$ such that 
$M = S \cdot S^{T}$. In general, this factorization cannot be done exactly since the left hand side typically has rank $D^2$, while the right hand side has rank at most $D$.  

However an approximate solution can be obtained as follows. The idea is to choose the matrix $S$ that minimizes the error, $|M - S \cdot S^{T}|^2$; the optimal $S$ can then be found using the singular value decomposition of $M$. In more detail: first one writes $M_{li,jk} = \sum_n s_n U_{li,n} V^*_{jk,n}$ (here, $s_n$ are the singular values and $U,V$ are unitary matrices). Second, one truncates the matrices $U_{li,n}, V_{jk,n}$ keeping only those columns corresponding to the largest $D$ singular values. The result are $D^2 \times D$ matrices 
$\tilde{U}_{li,n}, \tilde{V}_{jk,n}$. Finally, one sets $S_{lin} = \sqrt{s_n} 
\tilde{U}_{li,n}$. This gives the required factorization - provided that we  
adjust the phase ambiguity $U_{li,n} \rightarrow U_{li,n} e^{i\phi_n}, V_{jk,n} 
\rightarrow V_{jk,n} e^{-i\phi_n}$ appropriately. In practice, it is often more convenient to ignore the phase adjustment issue and set $S^A_{lin} = \sqrt{s_n} \tilde{U}_{li,n}$, $S^B_{jkn} = \sqrt{s_n} \tilde{V}^*_{jk,n}$. The result is a factorization $\sum_{n} S^A_{lin} S^B_{jkn} \approx \sum_{m} T_{ijm} T_{klm}$ where $S^A$ and $S^B$ differ by some phase factors. The TRG procedure can be applied as before - the only difference being that we have to keep track of two different tensors $T^A,T^B$ for the $A$ and $B$ sublattices. 

We will show in the next section that the error for this optimal decomposition is independent of the number of iterations and can be made arbitrarily small by increasing $D$. Indeed, the error vanishes as $\epsilon \sim \exp(-\text{const} \cdot (\log D)^2)$ - the same scaling behavior as the truncation error in DMRG. \cite{S0559} 

\begin{figure}[tb]
\centerline{
\includegraphics[width=2.0in]{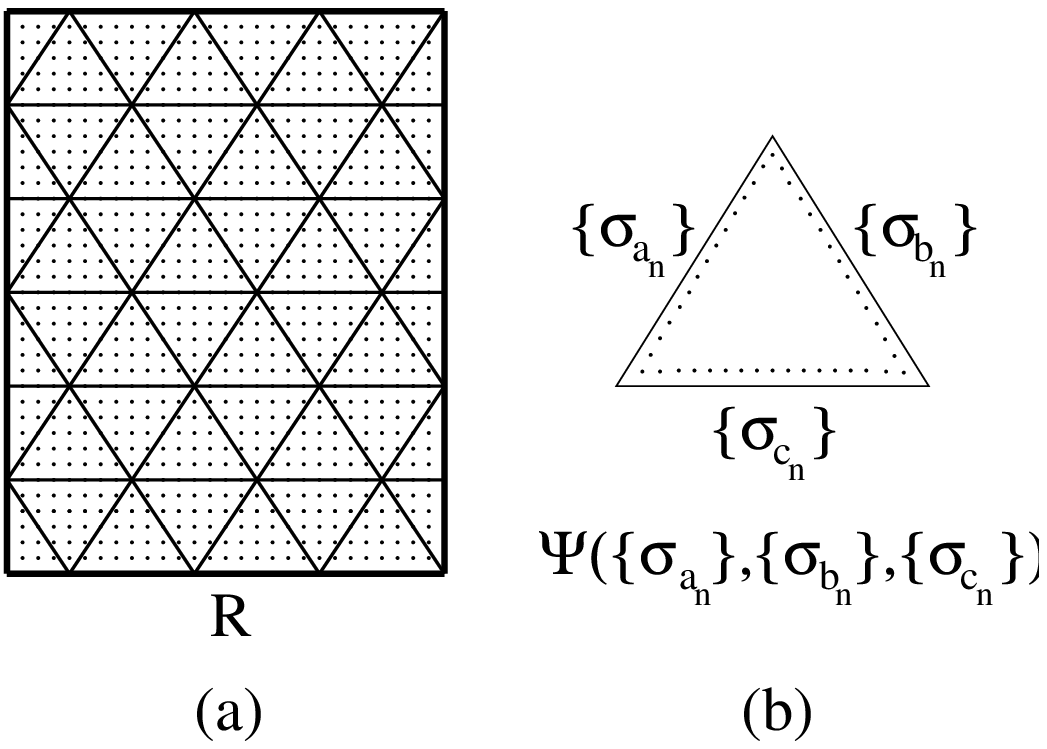}
}
\caption{
After (a) dividing $R$ into triangles, (b) the partition function of each triangle can be written as a function $\Psi(\{\si_{a_n}\},\{\si_{b_n}\},\{\si_{c_n}\})$ of the boundary spins.
}
\label{trilatt2}
\end{figure}

\textsl{Physical picture}:
In this section we explain the physics behind the TRG method, and give a 
physical interpretation of the fixed point tensor $T^*$. We begin by showing how arbitrary classical models in two dimensions can be thought of as tensor models on the honeycomb lattice. For concreteness we frame our discussion around the case of the square lattice Ising model, $Z = \sum_{\{\si_i\}} \exp({K\sum_{\<ij\>} \si_i\si_j})$. Consider the partition function $Z_R$ for some finite region $R$ in the plane. One way to compute $Z_R$ is to triangulate $R$, dividing it into triangles of size $L$ much larger 
than the lattice spacing $l$ (Fig. \ref{trilatt2}a). 

Consider one of the triangles. Imagine summing 
over all the lattice degrees of freedom within the triangle. The result will yield some number $\Psi= \Psi(\{\si_{i_n}\})$ that depends on the values of $\{\si_{i_n}\}$ at the 
boundary of the triangle (Fig. \ref{trilatt2}b). It is convenient to separate out these boundary degrees of freedom into three groups 
$\{\si_{a_n}\},\{\si_{b_n}\},\{\si_{c_n}\}$ corresponding to the three sides 
$a,b,c$ of the triangle. Denoting $\{\si_{a_n}\}$ schematically by $\alpha$, and 
similarly for $\{\si_{b_n}\},\{\si_{c_n}\}$, we can think of $\Psi$ as a 
three index tensor $\Psi = \Psi_{\alpha\beta\gamma}$. 

To obtain the partition function for the region $R$, we simply need to glue together 
all the triangles and then sum over the spins at their boundaries: $Z_R = 
\sum_{\alpha\beta\gamma\delta\epsilon...} 
\Psi_{\alpha\beta\gamma}\Psi_{\alpha\delta\epsilon}...$. This is nothing but 
the partition function of a tensor model (\ref{parttensor}).
(Actually, a more careful analysis shows that the resulting tensor model has different tensors $\Psi^A, \Psi^B$ for the $A,B$ sublattices).

\begin{figure}[tb]
\centerline{
\includegraphics[width=2.5in]{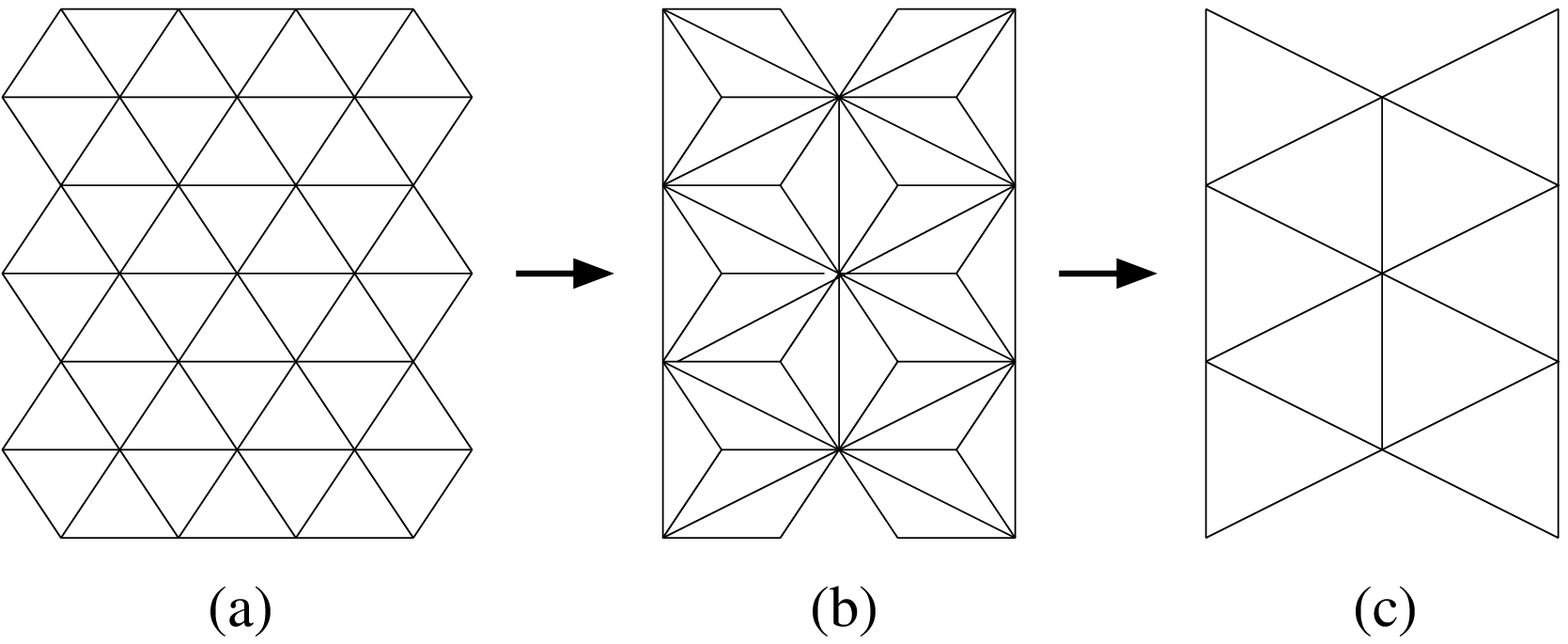}
}
\caption{
The tensor renormalization group transformation can be viewed as a two step
change in the triangulation of $R$.
}
\label{trirgsum}
\end{figure}

It is useful to think about the tensor renormalization group transformation in
this context. Recall that the first step is to find tensors 
$S$ satisfying $\sum_{n} S_{lin} S_{jkn} \approx \sum_{m} T_{ijm} T_{klm}$.
Thinking in terms of the triangles, the right hand side is simply 
the partition function of the rhombus obtained by gluing two triangles together. 
Thus, a solution $S$ can be constructed by setting $S$ equal to the partition 
function of one of the obtuse triangles obtained by dividing the rhombus in the other
direction:
\begin{align}
\includegraphics[height=0.4in]{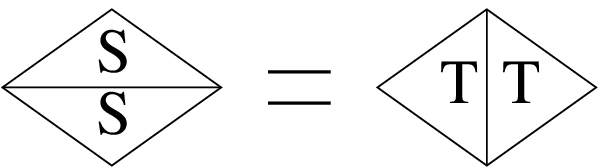} \label{trireconn}
\end{align}
Thus, the first step of the renormalization transformation process simply 
changes the triangulation as shown in Fig. \ref{trirgsum}b. The second step also has 
a simple interpretation. Examining the definition of $T'$, it is not hard to 
see that $T'$ is simply the partition function of a large equilateral 
triangle, obtained by gluing together three obtuse triangles:
\begin{align}
\includegraphics[height=0.4in]{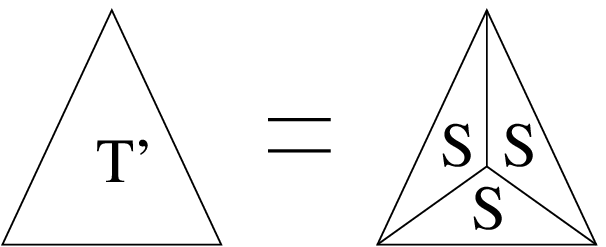} \label{tricoarse}
\end{align}
Thus, the second step simply glues together triplets of obtuse triangles
to form larger equilateral triangles as shown in Fig. \ref{trirgsum}c. In this way, the TRG 
method builds up larger and larger triangles.

However, there is one subtlety. As we build up 
larger and larger triangles, the corresponding tensors will have indices with 
larger and larger ranges, increasing from $2^{L/l}$ to $2^{\sqrt{3}L/l}$ to 
$2^{3L/l}$ and so on. Yet, the TRG method insists on approximating these tensors by a 
tensor with a fixed range $D$. How can this approximation possibly be accurate? 
 
To answer this question, we must explain the physical meaning of 
$\Psi_{\alpha\beta\gamma}$. Let $\Psi$ be a tensor obtained from the partition 
function of a very large triangle. Writing out the labels $\alpha,\beta,\gamma$ 
explicitly, we can write $\Psi$ as a function 
$\Psi(\{\si_{a_n}\},\{\si_{b_n}\},\{\si_{c_n}\})$ 
of the spins on the three sides of the triangle. In fact, this function should be 
thought of as a \emph{wave function} for a one dimensional quantum spin system with 
spins living on the boundary of a triangle. This interpretation comes from thinking of the original $2$ dimensional classical model (e.g. the Ising 
model) as a $(1+1)$ dimensional quantum model. We think of the 
direction parallel to the boundary of the triangle as space, and the (radial) 
direction perpendicular to the boundary as time. We imagine constructing a one 
dimensional transfer matrix/quantum Hamiltonian $H$ living on the 
boundary of the triangle, whose (radial) time evolution generates the two dimensional 
classical model in question. Then in this picture, the function  
$\Psi(\{\si_{a_n}\},\{\si_{b_n}\},\{\si_{c_n}\})$ is the result of evolving $H$ for 
a long time (the triangle is large). Hence $\Psi$ is simply the ground state 
of $H$ - up to exponentially small corrections.

If we assume that the original classical model (e.g. the Ising model) is not
critical, then $\Psi$ is the ground state of a \emph{gapped} Hamiltonian. 
Gapped ground states in one dimension have an important property: they are only 
weakly entangled. More specifically, it is known that the density matrix 
$\rho(x,L)$ of a region of size $x$ converges to a fixed density matrix $\rho_\infty$ 
as $x,L \rightarrow  \infty$. Moreover, the size of the $m$th eigenvalue $\lambda_m$ of 
$\rho_\infty$ falls off rapidly with increasing $m$: $\lambda_m \sim 
\exp(-\text{const} \cdot \log(m)^2)$. \cite{OHA9927} 

\begin{figure}[tb]
\centerline{
\includegraphics[width=2.8in]{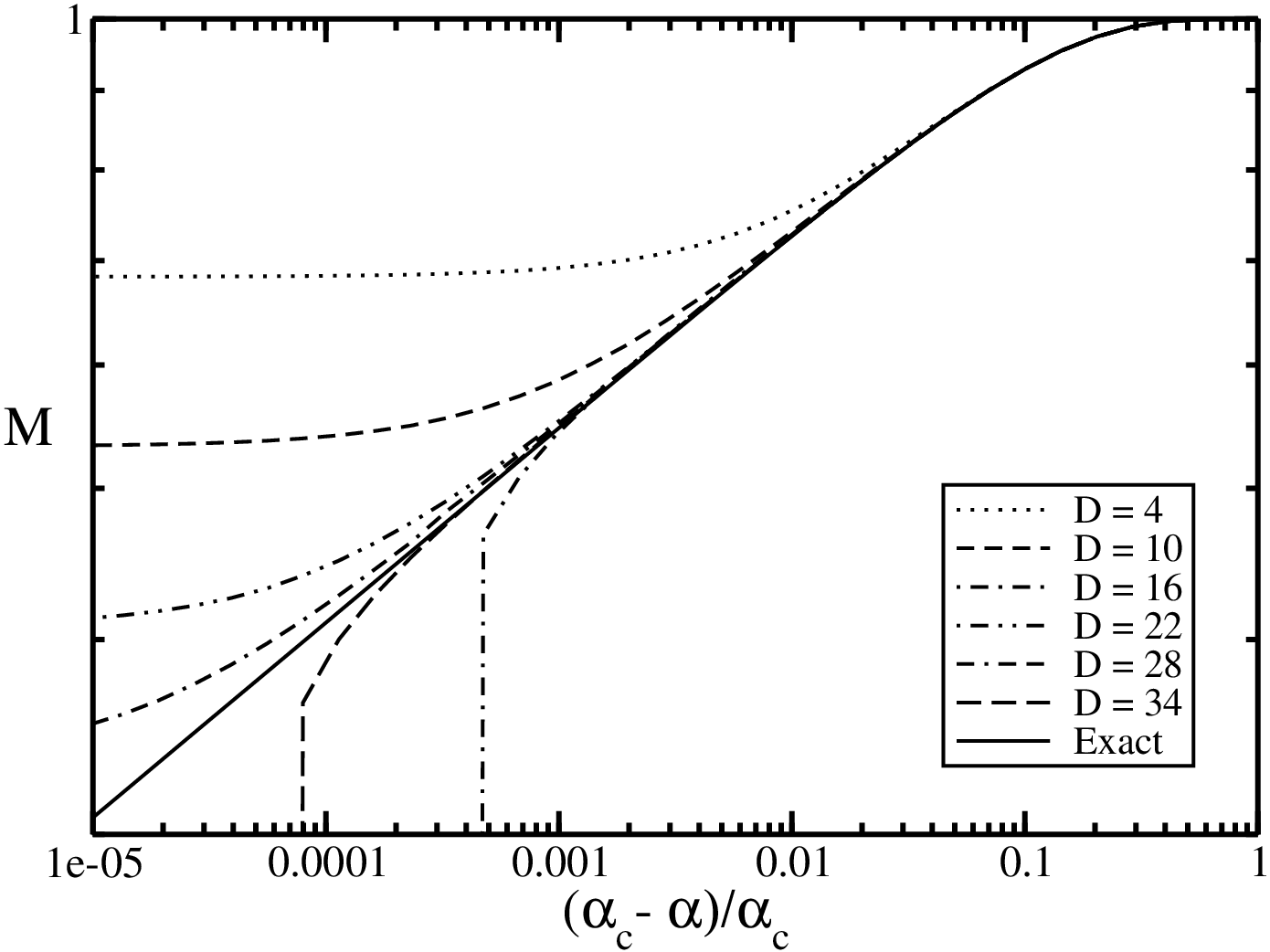}
}
\caption{
The magnetization of the triangular lattice Ising model as a function of $\alpha = e^{-2\beta J}$, obtained using the TRG method.  
}
\label{ferrosum}
\end{figure}


It is this property which guarantees the accuracy of the TRG method. Indeed, as a 
consequence of this property, one can factor $\Psi$ into a product of three spin 
states on the three sides of the triangle,
\begin{equation}
\Psi(\si_{a},\si_{b},\si_{c}) 
\approx \sum_{ijk=1}^{D} T_{ijk}  
\Psi^i_A(\si_{a})\Psi^j_B(\si_{b})\Psi^k_C(\si_{c})
\label{factor} 
\end{equation}
with high accuracy. Such a factorization can be obtained by 
choosing $\Psi^i_A,\Psi^i_B,\Psi^i_C$ to be the $i$th largest eigenstates of the 
density matrices of sides $A,B,C$, and letting $T_{ijk}$ be the matrix
elements of the tensor $\Psi(\si_{a},\si_{b},\si_{c})$ between these states. By 
the discussion above, the error $\epsilon$ of this representation is (1) independent 
of the size $L$ of the triangle, and (2) decreases rapidly with increasing $D$
($\epsilon \sim \exp(-\text{const} \cdot (\log D)^2)$). 

Thus, even though the exact tensor $\Psi_{\alpha\beta\gamma}$ has an exponentially large range, one can make a change of basis so that in that basis, 
$\Psi$ can be accurately approximated by a tensor $T_{ijk}$ whose indices have 
a fixed range $D$. The TRG method can be thought of as a numerical technique for 
(approximately) constructing this tensor $T_{ijk}$ for larger and larger triangles. 
The fixed point $T^*$ is the value of this tensor in the limit of an infinitely large 
triangle. 

The above analysis was based on the assumption that the classical model was not
critical. If instead the classical model is critical, the associated quantum states are \emph{gapless} ground states. Gapless 
ground states are more entangled then their gapped counterparts. The 
entanglement entropy $S = -\Tr(\rho \log \rho)$ of a region of size $x$ in a system of 
size $L$ grows logarithmically with the region size, $S \sim \log x$. \cite{VLR0302}
This means that the factorization (\ref{factor}) will always break down when the 
triangle is sufficiently large. Thus, in principle the TRG method - like DMRG \cite{S0559}-
breaks down at criticality. 

\textsl{A simple example}:
In this section we demonstrate the method with a simple example: the triangular lattice
Ising model: $Z = \sum_{\{\si\}} \exp(\beta J\sum_{\<ij\>} \sigma_i \sigma_j)$.
Note that the Ising model partition function can be written as a sum over domain wall configurations where the domain walls live on the bonds of
the honeycomb lattice. This domain wall model can be easily realized by a tensor network with $D = 2$. We think of the state $i=1$ as denoting "no domain wall" and $i=2$ as denoting 
"domain wall." Then the tensor with nonzero components
\begin{align}
T^{111} &= 1, \ T^{122} = T^{212} = T^{221} = \alpha,
\end{align}
$\alpha = e^{-2\beta J}$ gives rise to the correct Boltzmann weight.

Applying the TRG method to the above tensor, we compute the free energy per unit
spin, $F = -\frac{1}{N\beta}\log(Z)$ in the thermodynamic limit $N \rightarrow \infty$.
The magnetization $M$ can be obtained by taking numerical derivatives of $F$ (though we need to use a more complicated tensor $T$ to represent an Ising model with an external magnetic field $H$). Increasing $D$, the computation rapidly converges to the exact result \cite{H5025} except in an increasingly narrow interval around the critical point $\alpha_c = 1/\sqrt{3}$ (Fig. \ref{ferrosum}). 

This interval becomes so narrow that one can study the critical point itself. For example, the magnetization curve for $D = 34$ predicts an $\alpha_c$ within $10^{-4}$ of the exact result. One can even estimate the critical exponent 
$\beta$ from the scaling behavior of the magnetization. We find $\beta = 0.12$, not far from the exact value $\beta = 1/8$. However, as explained earlier, the TRG method (like DMRG) is best suited to studying systems \emph{off criticality}. An interesting question for further research is whether the TRG method can be modified so that it's (almost exponential) accuracy is uniform, both away from and near critical points.


The authors would like to thank Michael Freedman and Frank Verstraete for motivating this problem and Nihat Berker for useful discussions. This work was supported by the Harvard Society of Fellows (M.L.), and by NSF grant DMR-0517222 (C.P.N.).

\newcommand{\noopsort}[1]{} \newcommand{\printfirst}[2]{#1}
  \newcommand{\singleletter}[1]{#1} \newcommand{\switchargs}[2]{#2#1}

\end{document}